\documentclass[preprint,aps,12pt,showpacs,showkeys,nofootinbib,tightenlines]{revtex4}
\usepackage{epsfig}
\usepackage{subfigure}
\usepackage{graphicx,rotating,hyperref,slashed,amsmath,color,amssymb,amsfonts}
\usepackage{array,multirow}

\begin{document}

\title{750 GeV diphoton excess confronted with a top-pion \\
in the TTM model}

\author{Yu-Chen Guo$^1$ \footnote{Email: lgguoyuchen@126.com}, Chong-Xing Yue$^1$
and Zhen-hua Zhao$^{1,2}$}

\affiliation{$^1$Department of Physics, Liaoning Normal University, Dalian 116029,  China\\
$^2$Institute of High Energy Physics, Chinese Academy of Sciences, Beijing 100049, China
\vspace*{1cm} }

\begin{abstract}
The latest LHC data suggest an intriguing excess at $m^{}_{\gamma \gamma}=750$ GeV
which apparently requires an explanation from the beyond standard model physics.
In this note we explore the possibility for this signal to arise from a top-pion in
the Top Triangle Moose model which can be viewed as a dimensional-deconstruction
version of the top-color assisted technicolor model. We demonstrate that the observed excess
can be accommodated by and has important implications for this interesting model.
\end{abstract}

\pacs{12.60.Nz, 13.85.Qk, 14.80.Tt}
\keywords{TTM model, Diphoton excess}

\maketitle


\section{Introduction}

Recently, the first data obtained at the LHC Run 2 with a center-of-mass energy $\sqrt{s}=13$ TeV
have been released. Interestingly, both the ATLAS and CMS collaborations have observed an excess
in the diphoton invariant mass distribution around 750 GeV, with the 3.9$\sigma$ \cite{ATLAS:2015}
and 2.6$\sigma$ \cite{CMS:2015} significance respectively. Interpreted in terms of a resonance induced
by one (pseudo)scalar $S$ of the mass 750 GeV, this signal indicates that the production cross section
times the $\gamma \gamma$ branching ratio of $S$ has a value of \cite{ATLAS:2015,CMS:2015}
\begin{equation}
\sigma(pp \to S) {\rm Br}(S \to \gamma \gamma) =
\left\{
\begin{array}{lcl}
(10\pm 3) & \text{fb} & (\sqrt{s} = 13 \text{TeV ATLAS} ) \;,    \\
(6 \pm 3)  & \text{fb} &(\sqrt{s} = 13 \text{TeV  CMS} ) \;.
\end{array}
\right.
\label{1}
\end{equation}
In addition, the ATLAS excess features a width of about 45 GeV at the face value while the CMS excess width is
narrow and inconclusive \cite{1512.04933}. We will therefore take seriously the suggestive peak position
and signal cross section, but just consider the width value as a reference in the following analysis.

If established, this excess would very likely lead us to some new physics beyond the standard model (SM).
In this connection various interesting possibilities have been proposed to explain the signal \cite{75001,
75002,75003,75004,75005,75006,75007,75008,75009,75010,75011,75012,75013,75014,75015,75016,75017,75018,75019,
75020,75021,75022,75023,75024,75025,75026,75027,75028,75029,75030,75031,75032,75033,75034,75035,75036,75037,
75038,75039,75040,75041,75042,75043,75044,75045,75046,75047,75048,75049,75050,75051,75052,75053,75054,new1,
new2,new3,new4,new5,new6,new7,new8,new9}.
One solution of particular interest is to consider $S$ as a composite particle due to the confinement of a new strong dynamics \cite{75055,75056,75057,75058,75059,75060,75061,75062,75063,75064,75065,75066,75067,75068,75069}.
This idea can be traced back to the technicolor (TC) theory which serves as an alternative for the
electroweak symmetry breaking \cite{TC}. In TC models, there are generally many massless techniquarks
introduced which will result in a large chiral symmetry.
When the chiral symmetry is spontaneously broken by the techniquark condensation, some Goldstone bosons
corresponding to the broken symmetries (except those to the broken electroweak symmetries) will arise
which are unique composite pseudoscalar candidates. An interesting TC model is the one assisted by
an additional strong interaction --- top-color, as the name suggests, which only acts on
the third generation quarks \cite{topcolor}.
This model can naturally accounts for the largeness of $m^{}_t$ via the top-quark condensation,
while the electroweak symmetry breaking still mainly comes from the conventional TC sector.
Noteworthy, there will be three additional Goldstone bosons called top-pions that correspond to
the broken chiral symmetries involving $t$ and $b$.

In this short note we will investigate the possibility for $S$ to be a neutral top-pion from the
Top Triangle Moose (TTM) model \cite{TTM,TTM2,TTM3,TTM4,TTM5} which can be viewed as a dimensional-deconstruction \cite{DD1,DD2}
version of the top-color assisted TC model \cite{TC21,TC22,TC23,TC24,TC25}. In the next section, we will first recapitulate
the essential features and relevant details of TTM model and then perform numerical analysis
to show that the experimental result can be fitted.
Our conclusions and some further discussions will be given in the last section.

\section{ FRAMEWORK and Analysis }
\noindent

\subsection{The TTM model revisited}

The TTM model \cite{TTM,TTM1} was originally proposed for reducing the tension between obtaining a
realistic top-quark mass and fulfilling the electroweak precision measurements, which exists in the three-site
Higgsless model \cite{Three}. This is achieved by separating the bulk of electroweak symmetry
breaking from that of top-quark mass generation, resembling the top-color assisted TC
model \cite{topcolor}. In the TTM model, one invokes the gauge interactions
${\rm SU}(2)^{}_0 \times {\rm SU}(2)^{}_1 \times {\rm U}(1)^{}_2$ (with the respective couplings
$g$, $\tilde g$ and $g^\prime$) under which the fermion fields furnish the following representations:
\begin{eqnarray}
 Q^{}_{L0} \sim (2, 1, \frac{1}{6}) \;, \hspace{0.5cm}  Q^{}_{L1} \& Q^{}_{R1} \sim (1, 2, \frac{1}{6}) \;,
\hspace{0.5cm}  u^{}_{R2} \sim (1,1, \frac{2}{3}) \;, \hspace{0.5cm} d^{}_{R2} \sim (1,1, -\frac{1}{3}) \;.
\label{2}
\end{eqnarray}
Note that we introduce a vector-like partner for every SM fermion, which is allowed to assume
a mass term $M^{}_D \overline{Q^{}_{L1}} Q^{}_{R1}$ in the first place.

In order to break
${\rm SU}(2)^{}_0 \times {\rm SU}(2)^{}_1 \times {\rm U}(1)^{}_2$ down to the electromagnetism interaction
as well as generate masses for the massless quarks, two bi-fundamental
fields $\Sigma^{}_{01}$ and $\Sigma^{}_{12}$ together with a top-Higgs field $\Phi$ are introduced.
While $\Phi$ carries a quantum number of $(2,1,-1/2)$, $\Sigma^{}_{01}$ and $\Sigma^{}_{12}$ are
non-linear sigma model fields with the related covariant derivatives being
\begin{eqnarray}
D^{}_\mu \Sigma^{}_{01} &=& \partial^{}_\mu \Sigma^{}_{01} + i g W^{}_{0\mu} \Sigma^{}_{01}
-i \tilde{g} \Sigma^{}_{01} W^{}_{1\mu} \;, \nonumber \\
D^{}_\mu \Sigma^{}_{12} &=& \partial^{}_\mu \Sigma^{}_{12} + i \tilde{g} W^{}_{1\mu} \Sigma^{}_{12}
-i g^\prime \Sigma^{}_{12} \tau^3 B^{}_\mu \;.
\label{3}
\end{eqnarray}
It is understood that $W^{}_{0 \mu}= W^{a}_{0 \mu} \tau^a$ and
$W^{}_{1 \mu}= W^{a}_{1 \mu} \tau^a$ with $\tau^a=\sigma^a/2$ (for $a=1,2,3$) being the generators of
SU(2). These scalar fields will develop vacuum expectation values (VEVs) and can be expanded around
their VEVs as follows
\begin{eqnarray}
\begin{array}{l}  \vspace{0.2cm}
\Sigma^{}_{01} \sim \displaystyle \frac{1}{\sqrt2} v \cos{\omega} {\bf 1}^{}_{2 \times 2}
+ i \tau^a \pi^a_0 \;, \\
\Sigma^{}_{12} \sim \displaystyle \frac{1}{\sqrt2} v \cos{\omega} {\bf 1}^{}_{2 \times 2}
+ i \tau^a \pi^a_1 \;,
\end{array}  \hspace{1cm}
\Phi \sim \left( \begin{array}{c} \displaystyle \frac{1}{\sqrt2} \left(v \sin{\omega} + H^{}_t
+ i\pi^0_t \right) \\ i \pi^-_t  \end{array} \right) \;,
\label{4}
\end{eqnarray}
where $\omega$ is used to parameterize the partition of electroweak symmetry breaking (equivalently $v$) among
the different sectors. Six combinations of the above degrees of freedom will be eaten, generating
masses for the two sets of weak bosons --- $W^{\pm}$, $Z^0$ and $W^{\prime \pm}$, $Z^{\prime 0}$.
And there exists a mass relation between $W^{\pm}$ and $W^{\prime \pm}$ as given by
\begin{eqnarray}
\frac{M^2_W}{M^2_{W^\prime}}=\frac{g^2}{4 \tilde g^{ 2} \cos^2{\omega}} \;.
\label{5}
\end{eqnarray}
For phenomenological considerations (which demand $M^{}_{W} \ll M^{}_{W^\prime}$),
$x \equiv g/\tilde g$ is required to be a small quantity.
Hence we are left with four uneaten states, namely $H^{}_t$ (called top-Higgs) and one iso-triplet (called top-pions)
\begin{eqnarray}
\Pi^a_t= -\sin{\omega} \left( \frac{\pi^a_0 + \pi^a_1}{\sqrt 2} \right) + \cos{\omega} \ \pi^a_t \;.
\label{6}
\end{eqnarray}
They receive their masses mainly from the following interactions \cite{TTM,TTM1}:
\begin{eqnarray}
\lambda {\rm Tr} \left( M^\dagger M - \frac{f^2}{2} \right)^2 +
\kappa f^2 {\rm Tr} \left| M - \frac{f}{\sqrt 2} \Sigma^{}_{01} \Sigma^{}_{12} \right|^2 \;,
\label{7}
\end{eqnarray}
with
\begin{eqnarray}
M= \left( \begin{matrix}
i \pi^+_t & (f+ H^{}_t + i \pi^0_t )/ \sqrt 2 \cr
(f+ H^{}_t - i \pi^0_t )/ \sqrt 2 & i \pi^-_t
\end{matrix} \right) \;.
\label{8}
\end{eqnarray}
A straightforward calculation gives us the top-Higgs and top-pion masses as
\begin{eqnarray}
M^2_H =  2 v^2 ( \kappa + 4 \lambda) \sin^2{\omega} \;, \hspace{0.5cm}
M^2_{\Pi^{}_t} = 2 v^2 \kappa \tan^2{\omega} \;.
\label{9}
\end{eqnarray}
In addition, these interactions give rise to the trilinear coupling among three top-pions
\begin{eqnarray}
g^{}_{\Pi^0_t \Pi^+_t \Pi^-_t} \Pi^0_t \Pi^+_t \Pi^-_t =
2i\kappa v\sin^2\omega \tan^2\omega \Pi^0_t \Pi^+_t \Pi^-_t \;,
\label{10}
\end{eqnarray}
which will contribute to the loop process of diphoton emission.

On the other hand, $\Sigma^{}_{01}$,
$\Sigma^{}_{12}$ and $\Phi$ also contribute to fermion masses by virtue of the Yukawa interactions
\begin{eqnarray}
M^{}_D \left[ \epsilon^{}_L \overline{ Q^{}_{L0}} \Sigma^{}_{01} Q^{}_{R1}
+ \overline{ Q^{}_{L1} } \Sigma^{}_{12}
\left(\begin{array}{cc} \epsilon^{}_{uR} & \\ & \epsilon^{}_{dR}   \end{array} \right)
\left( \begin{array}{c} u^{}_{R2} \\ d^{}_{R2} \end{array} \right) \right]
+ \lambda^{}_u \overline{ Q^{}_{L0} } \Phi u^{}_{R2} +
\lambda^{}_d \overline{ Q^{}_{L0} } \Phi d^{}_{R2} \;.
\label{11}
\end{eqnarray}
Here $\epsilon^{}_L$ is a flavor-universal parameter and has to be approximately $x/\sqrt{2}$ in order for
the SM fermions to decouple from $W^\prime$ \cite{ID}, while $\epsilon^{}_{fR}$ is flavor dependent.
It should be mentioned that $\epsilon^{}_{tR}$ is subject to the precision measurement constraint \cite{Three}
\begin{eqnarray}
\frac{M^2_D \epsilon^4_{tR} }{16 \pi^2 v^2} < \mathcal O(10^{-3}) \;.
\label{12}
\end{eqnarray}
To mimic the role of top-color, $\Phi$ is assumed to couple with the third generation fermions
preferentially, implying that only $\lambda_t$, $\lambda_b$ and $\lambda_{\tau}$ of the $\lambda_f$'s are considerable.
The mass matrix for $t$ and its vector-like partner $T$ (similarly for $b$ and its vector-like partner $B$)
as well as the resulting mass eigenvalues thus turns out to be
\begin{eqnarray}
M^{}_t= M^{}_D \left( \begin{array}{cc}
a & \epsilon^{}_L \\  \epsilon^{}_{tR} & 1
\end{array}  \right) \hspace{0.3cm} \Longrightarrow \hspace{0.3cm}
\left\{ \begin{array}{l}
m^{}_t= \displaystyle \frac{1}{\sqrt2} \lambda^{}_t v \sin{\omega} \left[ 1 +
\displaystyle \frac{a \epsilon^2_L + a \epsilon^2_{tR}
+ 2 \epsilon^{}_L \epsilon^{}_{tR} } {2(-a + a^3)} \right] \;, \\
m^{}_T= M^{}_D \left[ 1 - \displaystyle \frac{ \epsilon^2_L +  \epsilon^2_{tR}
+ 2 a \epsilon^{}_L \epsilon^{}_{tR} } {2(-1 + a^2) } \right] \;,
\end{array} \right.
\label{13}
\end{eqnarray}
where $a$ stands for $\lambda^{}_t v \sin{\omega} / (\sqrt{2} M^{}_D)$.
For the mass matrices of the first two generation fermions, there will not be a term like the
$a$ in Eq. (\ref{13}). Consequently, the mass eigenvalues for these SM fermions and their vector-like partners
are about $M^{}_D \epsilon^{}_L \epsilon^{}_{fR}$ and $M^{}_D$, respectively.

\subsection{Numerical results}

Before performing the numerical calculations, let us present some relevant details and useful formulas
concerning the decay of $\Pi^0_t$. By inserting Eqs. (\ref{4}, \ref{6}) into Eq. (\ref{11}),
one may easily derive the couplings between $\Pi^0_t$ and the SM fermions \cite{TTM,TTM1}
\begin{eqnarray}
\frac{i \Pi^0_t}{v} \left( m^{}_t \cot{\omega} \overline{t^{}_L} t^{}_R + m^{}_b \cot{\omega} \overline{b^{}_L} b^{}_R
+ m^{}_\tau \cot{\omega} \overline{\tau^{}_L} \tau^{}_R +  m^{}_c \tan{\omega}
\overline {c^{}_L} c^{}_R  \right)  \;.
\label{14}
\end{eqnarray}
The decay widths for these processes can be calculated according to
\begin{eqnarray}
\Gamma(\Pi_{t}^{0} \rightarrow t \bar{t} )  =  \frac{3 m_{t}^2}{8\pi v^2} \cot^{2}{\omega} \; m_{\Pi_{t}} \beta_{t t} \;,
\hspace{1cm}
\Gamma(\Pi_{t}^{0} \rightarrow c \bar{c}) =  \frac{3 m_{c}^2}{8\pi v^2} \tan^{2}{\omega} \; m_{\Pi_{t}} \beta_{c c} \;,
\label{15}
\end{eqnarray}
where $\beta^{}_{\alpha \beta}$ reads as
\begin{eqnarray}
\beta_{\alpha \beta}  =  \lambda^{1/2} (\frac{m_\alpha^2}{m^2_{\Pi_t} }, \frac{m_\beta^2}{m^2_{\Pi_t} }) \;,
\hspace{0.5cm} {\rm with} \hspace{0.5cm}
\lambda(\alpha, \beta) = 1 + \alpha^2 + \beta^2 - 2\alpha - 2\beta - 2\alpha \beta \;.
\label{16}
\end{eqnarray}
Apparently, the width of $\Pi^0_t \to c \bar c$ may be comparable to
and even exceed that of $\Pi^0_t \to t \bar t$ when $\tan{\omega}$ is large enough,
while the widths of $\Pi^0_t \to b \bar b$ and $\Pi^0_t \to \tau \bar \tau$ are
always much smaller than that of $\Pi^0_t \to t \bar t$.
The decays to $t \bar T$ and $T \bar T$ will also be included, if kinematically allowed,
for which the associated couplings are obtained as \cite{TTM,TTM1}
\begin{eqnarray}
g^{}_{\Pi_{t}^0 \bar{T}_L T_R} & = & \frac{i\lambda_t}{4\sqrt{2}} \left[ \frac{ 2\sqrt{2} x \epsilon_{tR}
(a^2 + \cos{2\omega}) + (2a^2 \cos{\omega} + (a^2-1)\sin{\omega}\tan{\omega})(x^2+2 \epsilon_{tR}^2)}
{ (a^2-1)^2} \right] \;,   \nonumber \\
g^{}_{\Pi_{t}^0 \bar{t}_L T_R} & = & \frac{i\lambda_t} {(a^2-1)} \left[\frac{x \sec{\omega} (-1+3a^2+(1+a^2)
\cos{2\omega})}{8a} + \frac{\epsilon_{tR}\cos{\omega} }{\sqrt{2}} \right] \;,  \nonumber \\
g^{}_{\Pi_{t}^0 \bar{T}_L t_R} & = & \frac{i\lambda_t} {(a^2-1)} \left [\frac{x \cos{\omega} }{2} + \frac{\epsilon_{tR}
\sec {\omega} (-1+3a^2+(1+a^2) \cos{2\omega}) } {4\sqrt{2}a} \right] \;.
\label{17}
\end{eqnarray}

Similar to the SM Higgs $h$, $\Pi^0_t$ can also decay to two photons (gluons) through the charged (colored) fermion triangle
loops. And the corresponding decay widths are given by
\begin{eqnarray}
 \Gamma(\Pi_t^0 \rightarrow \gamma\gamma) & = & \frac{ G_\mu \alpha^{2} m_{\Pi^{}_t}^3} {128 \sqrt{2} \pi^3}
\left| \sum^{}_{f} N_{c} Q^{2}_{f} F(\tau)^A_{1/2} \alpha_{f}+
\frac{v g_{\Pi^0_t \Pi^+_t \Pi^-_t}}{2 M^2_{\Pi^{}_t}}F(\tau)^A_0 \right|^2 \;,  \nonumber \\
\Gamma(\Pi_t^0 \rightarrow gg) & = & \frac{G_\mu \alpha_{s}^{2} m_{\Pi^{}_t}^3} {36\sqrt{2}\pi^3}
\left| \frac{3}{4}\sum^{}_{f} F(\tau)^A_{1/2} \alpha_{f} \right|^2 \;,
\label{18}
\end{eqnarray}
by analogy with those for $h$, with $\alpha^{}_f \equiv g^{}_{\Pi^0_t \bar f f}/g^{}_{h \bar f f}$
being the relative coupling strength (e.g., $\alpha^{}_t=\cot{\omega}$).
In the above $G^{}_\mu$ stands for the Fermi constant $1.17 \times 10^{-5}$ GeV$^{-2}$, whereas the symbols $\alpha$,
$\alpha^{}_s$, $N^{}_c$ and $Q^{}_f$ are self-explanatory.
Furthermore, the loop functions $F(\tau)^A_{1/2}$ and $F(\tau)^A_{0}$
are given by \cite{Gunion}
\begin{eqnarray}
F(\tau)^A_{1/2}=2\tau f(\tau) \;, \hspace{0.5cm} F(\tau)^A_{0}=-[\frac{1}{\tau}-f(\tau)]\tau^2 \;,
\label{19}
\end{eqnarray}
with $\tau_f \equiv 4 m_f^2/M_{\Pi^{}_t}^2$ and
\begin{equation}
f(\tau) =
\begin{cases}
\arcsin^{2}\sqrt{1/\tau} & \text{if } \tau \geq 1 \;, \\
-\displaystyle \frac{1}{4}\left[\textrm{ln}(\eta_+/\eta_-)-i\pi  \right]^2  & \text{if } \tau < 1 \;,
\end{cases}
\label{20}
\end{equation}
for $\eta_{\pm}=(1\pm\sqrt{1-\tau})$. In the summation over fermions, we will include $t$, $b$, $c$, $T$ and $\tau$.
In contrast, only $\Pi^{\pm}_t$ take effect via the scalar loop.

In order to reproduce the observed diphoton excess in terms of the $\Pi^0_t$ resonance, one needs to have
\begin{equation}
\sigma( pp \to \Pi_t^0 \to \gamma\gamma) =\frac{1}{ m^{}_{\Pi^{}_t} \Gamma s}
\left[ C_{gg} \Gamma(\Pi_t^0 \to gg) + \sum^{}_q C^{}_{q \bar q} \Gamma(\Pi^0_t \to q \bar q) \right]
\Gamma(\Pi_t^0\to \gamma\gamma)
\label{21}
\end{equation}
$(10 \pm 3) {\rm fb}$ for $\sqrt{s}=13$ TeV. Here $\Gamma$ is the total width of $\Pi^0_t$ and approximates to
\begin{eqnarray}
\Gamma \simeq & & \Gamma(\Pi_{t}^{0} \rightarrow gg)
+\Gamma(\Pi_{t}^{0}\rightarrow t\bar{t})+\Gamma(\Pi_{t}^{0}\rightarrow b\bar{b})
+\Gamma(\Pi_{t}^{0}\rightarrow c \bar{c})
+\Gamma(\Pi_{t}^{0}\rightarrow \tau \bar{\tau}) \nonumber \\
& & +\Gamma(\Pi_{t}^{0}\rightarrow \bar{t}T)
+\Gamma(\Pi_{t}^{0}\rightarrow T \bar{T}) \;,
\label{22}
\end{eqnarray}
while $C^{}_{gg}$ and $C^{}_{q \bar q}$ are the dimensionless partonic integral functions \cite{1512.04933}
\begin{eqnarray}
C_{gg} &=& \frac{\pi^2 }{8} \int_{m^2_{\Pi^{}_t}/s}^1 \frac{dz}{z} g(z) g(\frac{M^2}{sz}) \;, \nonumber \\
C^{}_{q \bar q} &=& \frac{4 \pi^2}{9} \int^{1}_{m^2_{\Pi^{}_t}/s} \frac{dz}{z} \left[
q(z) \bar{q} (\frac{m^2_{\Pi_t}}{sz}) + \bar{q}(z) q (\frac{m^2_{\Pi^{}_t}} {sz} ) \right] \;.
\label{23}
\end{eqnarray}

%
\begin{figure}[htb]
\begin{center}
\includegraphics [scale=1.3] {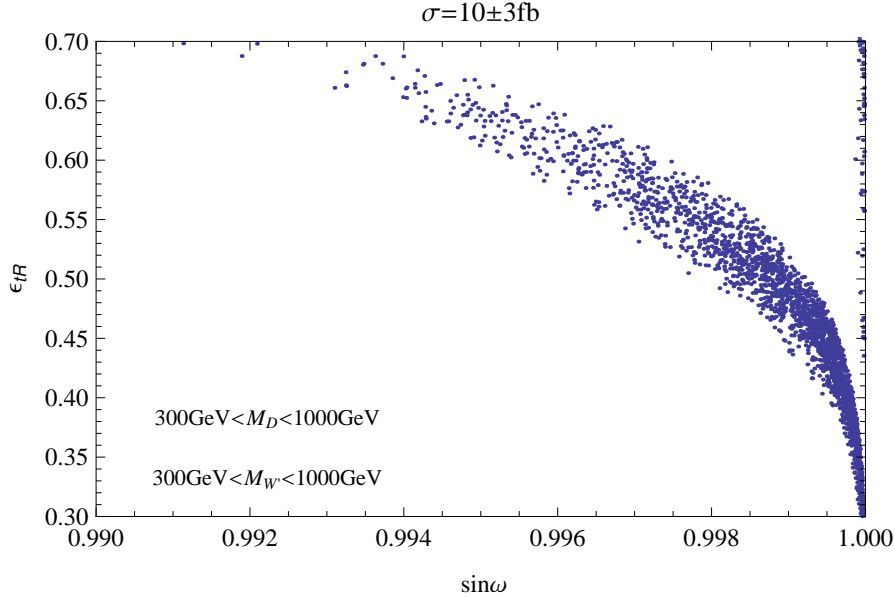}
\caption{The allowed parameter space of $\sin{\omega}$ versus $\epsilon^{}_{tR}$. }
\end{center}
\end{figure}
%
\begin{figure}[htb]
\begin{center}
\includegraphics [scale=0.77] {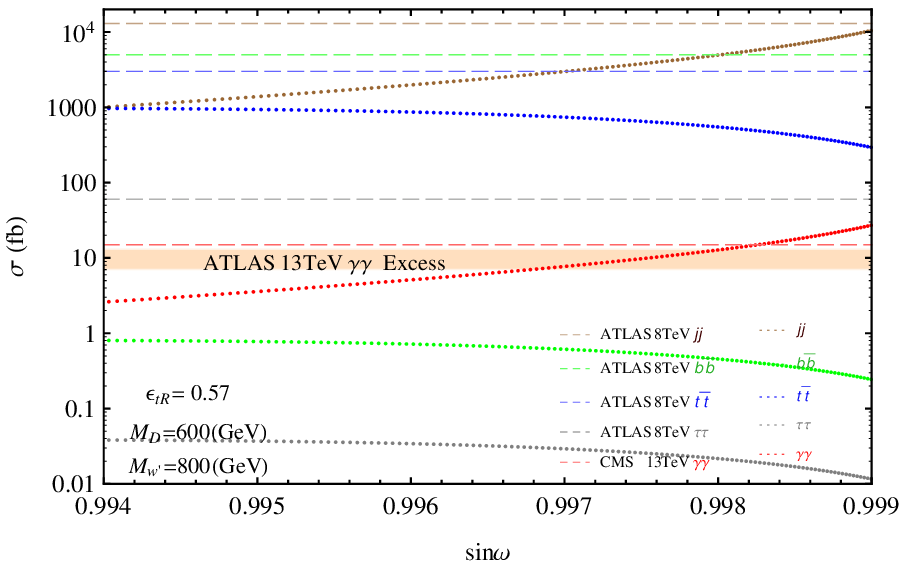}
\includegraphics [scale=0.75] {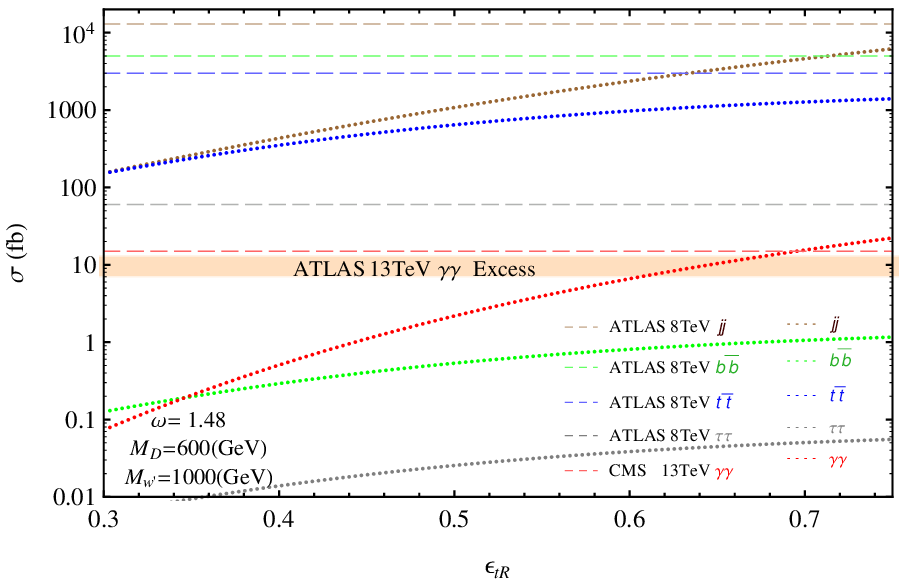}
\includegraphics [scale=0.75] {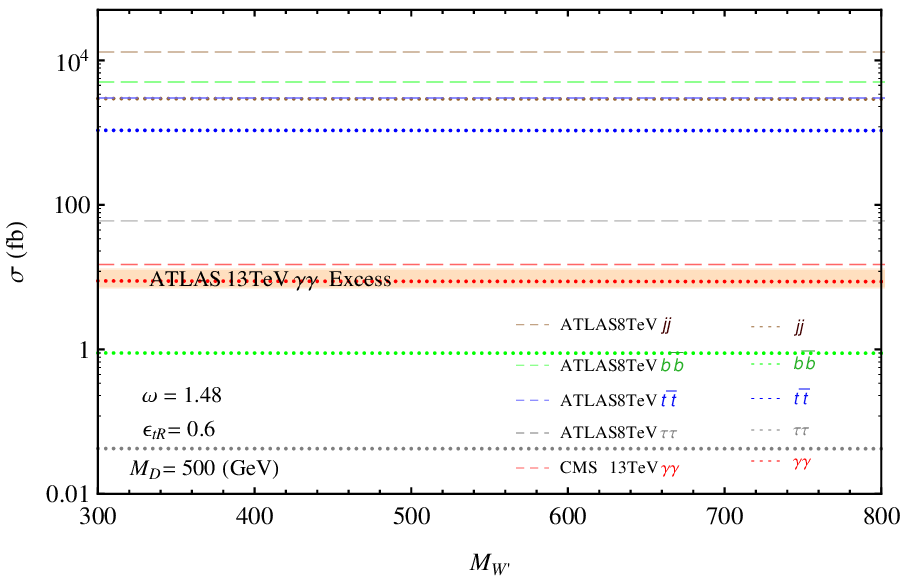} \hspace{0.1cm}
\includegraphics [scale=0.75] {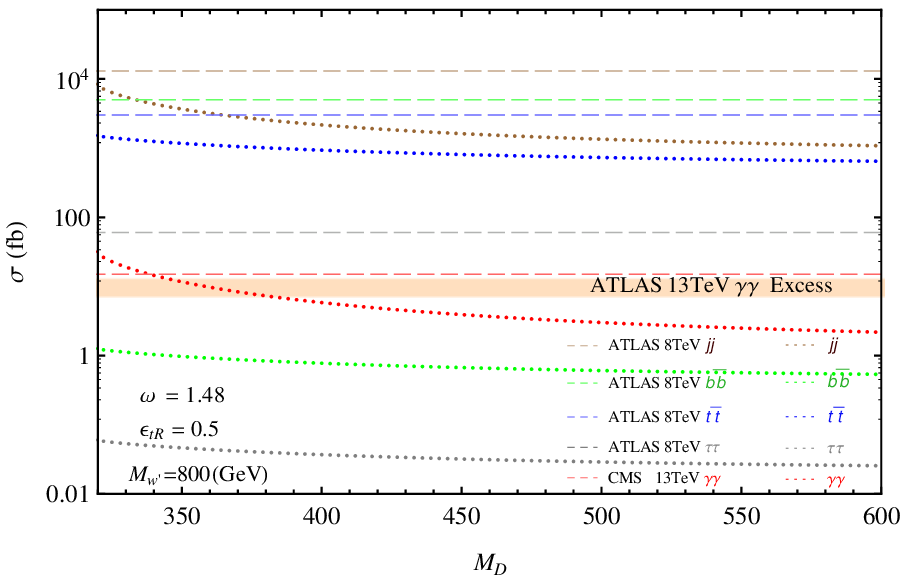}
\caption{The cross sections of $pp \to \Pi^0_t \to jj,\gamma \gamma,
t \bar t, b \bar b, \tau\tau$ (the $jj$ here include $gg$ and $cc$) at $\sqrt{s}=13$ TeV against the
four free parameters. The upper bounds for these cross sections
extrapolated from the LHC results at $\sqrt{s}=8$ TeV based on the approximation $\sigma_{13TeV}/\sigma_{8TeV}\approx 5$
have also been shown. }
\end{center}
\end{figure}

Now, we proceed to study whether the neutral top-pion $\Pi^0_t$ can serve as a candidate for the
observed resonance at 750 GeV. First of all, we will find the parameter space for $\Pi^0_t$ to give a signal
given by Eq. (\ref{21}). There are totally four free parameters --- $\sin{\omega}$,
$\epsilon^{}_{tR}$, $M^{}_{W^\prime}$ and $M^{}_D$ --- relevant for our study. The allowed parameter
space of $\sin{\omega}$ and $\epsilon^{}_{tR}$ is shown in Fig. 1. In this calculation, we have
imposed the constraint in Eq. (\ref{12}) and allowed $M^{}_{W^\prime}$ and $M^{}_D$ to
vary in the range [300, 1000] GeV. As one can see from Fig. 1, $\sin{\omega}$
has to lie in a range extremely close to 1 while $\epsilon^{}_{tR}$ ranges from 0.3 to 0.7.
In Fig. 2 we have presented the cross sections of $ pp \to \Pi^0_t \to
t \bar t, b \bar b, \tau\tau, gg, \gamma \gamma$ against one of the four parameters when the other
three parameters are fixed to some particular values.
It is apparent that these cross sections are more sensitive to $\sin{\omega}$ and $\epsilon^{}_{tR}$
than the mass parameters $M^{}_{W^\prime}$ and $M^{}_D$.
In particular, the cross section for $ pp \to \Pi^0_t \to \gamma \gamma$
can be enhanced by increasing $\sin{\omega}$ and $\epsilon^{}_{tR}$.
The upper limits for the cross section of each decay mode are also shown, which
are extrapolated from the LHC searches at $\sqrt{s}=8$ TeV \cite{tt8,bb8,gg8a,gg8b}.
One can see that the phenomenological consequences of a 750 GeV $\Pi^0_t$
are compatible with these constraints.
Finally, we plot the possible total decay width of $\Pi^0_t$ as a function of $\sin{\omega}$ in Fig. 3.
In most of the allowed parameter space, $\Gamma$ is within 30 GeV, in agreement the observed one.
But as $\sin{\omega}$ decreases from 1 $\Gamma$ will grow rapidly,
rendering the signal for $\sigma(gg \to \Pi^0_t\to \gamma \gamma)$ unobservable.

Finally, let us briefly discuss the properties of $\Pi^{\pm}_t$ and $H^{}_t$.
An interesting consequence of the TTM model is that it predicts
$\Pi^{\pm}_t$ to have the same mass as $\Pi^{0}_t$ (i.e., 750 GeV).
On the other hand, the corresponding couplings for $\Pi^{\pm}_t$ are given by
\begin{eqnarray}
\frac{i \sqrt{2} \Pi^+_t}{v} \left( m^{}_t \cot{\omega} \overline{t^{}_R} b^{}_L + m^{}_b \cot{\omega} \overline{t^{}_L} b^{}_R
+ m^{}_\tau \cot{\omega} \overline{ \nu^{}_{\tau L} } \tau^{}_R +  m^{}_c \tan{\omega}
\overline {c^{}_R} s^{}_L  \right) + h.c. \;.
\label{24}
\end{eqnarray}
One immediately finds that the dominant decay channel of $\Pi^{+}_t$ will be $\Pi^{+}_t \to t \bar b$ or $\Pi^+_t \to
c \bar s$, depending on the value of $\sin{\omega}$. Accordingly, $\Pi^{+}_t$ can mainly be produced in association with a
quark via the process $g b \to b \to \bar t \Pi^+_t$ or $g s \to s \to \bar c \Pi^+_t$, while the contributions from electroweak processes such as $u \bar d \to W^+ \to H^{}_t \Pi^{+}_t$ are negligibly small. By comparing Fig. 10 and Fig. 13
of Ref. \cite{TTM1}, we can see that the production cross section for $\Pi^{\pm}_t$ is about one order of
magnitude smaller than that for $\Pi^0_t$. In combination with the fact that the dominant decay channels of $\Pi^{\pm}_t$
are hadronic which tend to be masked by high backgrounds, this means that a direct search for $\Pi^{\pm}_t$ is difficult
at the present stage.
Last but not least, we point out that $H^{}_t$ is a unique candidate
for the 125 GeV resonance discovered in 2012 \cite{125a,125b} which is widely believed to be the (SM) Higgs boson.
The reason lies in the fact that in the preferred parameter space region $\sin{\omega} \sim 1$
the electroweak symmetry breaking mainly originates from a non-vanishing VEV of the top-Higgs field $\Phi$,
leading $H^{}_t$ to behave in a similar way as the SM Higgs boson. The latter point can be easily seen
from the coupling strengths between $H^{}_t$ and the SM particles.
For instance, the coupling strength for $H^{}_t W^{+} W^{-}$ is approximately $g^2 v \sin{\omega}/2$ \cite{TTM,TTM1}
which can be reduced to that in the SM case by taking $\sin{\omega}=1$. In addition, from Eq. (\ref{9}) one can see
that $H^{}_t$ has a mass suppressed by $\cos{\omega}$ (which takes a value $\sim$0.1 in the viable parameter space region)
as compared to $M^{}_{\Pi^{}_t}=750$ GeV, in good qualitative agreement with the experimental results.
Quantitatively, one always has the freedom of specifying a suitable value for the $\lambda$ in Eq. (\ref{9})
to obtain an 125 GeV $H^{}_t$ precisely.

%
\begin{figure}[htb]
\begin{center}
\includegraphics [scale=1.3] {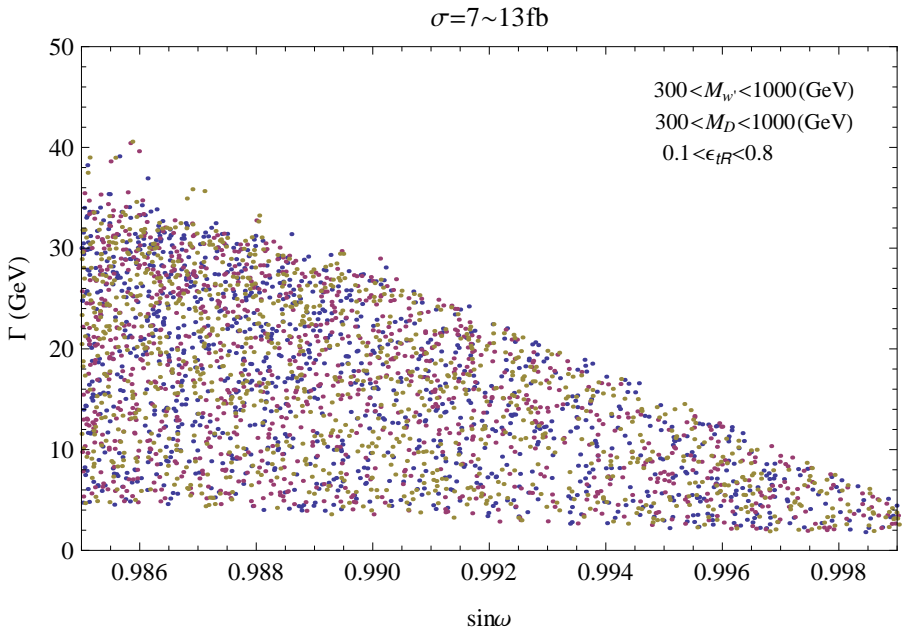}
\caption{The possible total width of $\Pi^0_t$ against $\sin{\omega}$.}
\end{center}
\end{figure}

\section{Summary }
\noindent

While the 750 GeV diphoton excess necessitates a further collection of data to be confirmed or ruled out,
it is still worthwhile for us to explore various new physics that can naturally account for this
signal and their physical implications. Motivated by the possibility that this excess may be due to
one composite (pseudo)scalar of a mass around 750 GeV, we investigate whether TC models can
accommodate such a signal and find that a variety of TC models are unable to fit with the experimental result.
These specific models include the original one-family model of FS \cite{FS}, variant one family model \cite{one},
LR multiscale model \cite{LR}, TC Straw Man model low scale \cite{TCSW}, MR Isotriplet model \cite{MR} and top-color assisted
TC models \cite{topcolor}. Most of these models have a too large width for $\Pi^0_t \to t\bar{t}$ to
give a satisfying result. Only in variant one family model can the total width of $\Pi^0_t$ may be
lowered to $\sim 45$ GeV. But at this time the cross section is smaller that the observed signal
by two order of magnitudes.

Most of our attention has been paid to the TTM model which is phenomenologically viable.
It is found that the identification of $\Pi^0_t$ as the reported
resonance has important implications: The parameter space
(particularly that for $\sin{\omega}$) is strongly constrained, making the model more predictive.
Hence we may confirm or rule out the possibility of $\Pi^0_t$ being the 750 GeV resonance by
paying particular attention to this channel. Furthermore, the TTM model predicts two charged
top-pions $\Pi^{\pm}_t$ of the same mass as $\Pi^0_t$. Unfortunately, they are
difficult to discover owing to a combination of low production cross section and high
backgrounds for their decay products \cite{TTM1}. Finally, it is interesting to note that
$H^{}_t$ may serve as a unique candidate for the observed 125 GeV resonance.
To summarize, the TTM model can explain the recently
observed diphoton excess and has testable predictions, thereby deserving some attention.

\section*{Acknowledgements}

\noindent
One of us (Z. H. Z.) is indebted to Dr. Jue Zhang for helpful discussions and Prof.
Zhi-zhong Xing for reading the manuscript.
This work is supported in part by the National Natural Science Foundation of China
under grant No. 11275088 (Y. C. G. and C. X. Y.), grant No. 11135009 and
grant No. 11375207 (Z. H. Z); and by the Natural Science Foundation of the Liaoning Scientific
Committee under grant No. 2014020151.

\end{document}